**In-memory eigenvector computation in time $O(1)$**

*Zhong Sun\*, Giacomo Pedretti, Elia Ambrosi, Alessandro Bricalli, and Daniele Ielmini\**

Dr. Z. Sun, G. Pedretti, E. Ambrosi, Dr. A. Bricalli, Prof. D. Ielmini
Dipartimento di Elettronica, Informazione e Bioingegneria, Politecnico di Milano and IU.NET, Piazza L. da Vinci 32 – 20133 Milano, Italy
E-mail: zhong.sun@polimi.it; daniele.ielmini@polimi.it

Keywords: in-memory computing, eigenvector, time complexity, resistive memory, $O(1)$

**In-memory computing with crosspoint resistive memory arrays has gained enormous attention to accelerate the matrix-vector multiplication in the computation of data-centric applications. By combining a crosspoint array and feedback amplifiers, it is possible to compute matrix eigenvectors in one step without algorithmic iterations. In this work, time complexity of the eigenvector computation is investigated, based on the feedback analysis of the crosspoint circuit. The results show that the computing time of the circuit is determined by the mismatch degree of the eigenvalues implemented in the circuit, which controls the rising speed of output voltages. For a dataset of random matrices, the time for computing the dominant eigenvector in the circuit is constant for various matrix sizes, namely the time complexity is $O(1)$. The $O(1)$ time complexity is also supported by simulations of PageRank of real-world datasets. This work paves the way for fast, energy-efficient accelerators for eigenvector computation in a wide range of practical applications.**

Crosspoint resistive memory arrays have been intensively utilized to accelerate the matrix-vector multiplication (MVM),[1] which is an elementary operation in several algebraic problems, for instance, the training and inference of neural networks,[2,3] signal and image processing,[4,5] and the iterative solution of linear systems[6] or differential equations.[7] In such implementations, the crosspoint MVM is executed for several iteration cycles according to the algorithmic workflow, which might raise an issue in terms of processing time and energy efficiency of the computation. Recently, a crosspoint memory circuit architecture has been proposed and demonstrated for solving matrix equations in one step, including solving linear systems and computing eigenvectors.[8] Although the one-step solution capability can solve the inefficiencies of the iterative approach, the underlying time complexity of the circuit needs to be rigorously evaluated to assess the computing performance.

Eigenvector calculation is a fundamental problem in a broad scope of computing scenarios, *e.g.* webpage ranking,[9] facial recognition,[10] dynamic analysis and solving differential equations in fields such as physics and chemistry.[11] In the conventional computing paradigm, the dominant eigenvector (the eigenvector corresponding to the largest eigenvalue) of a matrix can be calculated using the power iteration method with a time complexity of $O(kN^2)$, where $N$ is the matrix size and $k$ is the number of iterations.[12] In this work, we show that the time complexity of the crosspoint memory circuit for computing eigenvectors is $O(1)$, namely the time to calculate a matrix eigenvector does not depend on the matrix size. Based on the feedback theory of operational amplifiers, we develop a numerical model to analyze the time response of the eigenvector circuit. Our time-dependent simulation results show that the time to calculate an eigenvector of an $N \times N$ matrix does not explicitly depend on $N$, *i.e.*, the time complexity is $O(1)$ for our circuit. We find that the computational time is governed by the highest eigenvalue of an associated matrix, which in turn is controlled by the mismatch degree between the practical and the nominal conductance values which implement the eigenvalue in the circuit. The $O(1)$ time complexity is further supported by circuit simulations of the calculation of dominant eigenvectors of random matrices and of PageRank evaluation for the Harvard500 database and its subsets.

Computing an eigenvector means solving the matrix equation

$$\boldsymbol{A}\boldsymbol{x} = \lambda \boldsymbol{x}, \tag{1}$$

where $\boldsymbol{A}$ is a square matrix, $\lambda$ is an eigenvalue of $\boldsymbol{A}$, $\boldsymbol{x}$ is the unknown eigenvector corresponding to $\lambda$. To solve Equation 1, matrix $\boldsymbol{A}$ is mapped by the conductance matrix $\boldsymbol{G_A}$ of a crosspoint memory array, which plays the role of a feedback network in a circuit (**Figure 1**a). The feedback configuration is enabled by transimpedance amplifiers (TIAs) and analog inverters. The conductance $G_\lambda$ of feedback resistors of TIAs represents eigenvalue $\lambda$. Since an analog eigenvalue cannot be exactly implemented, the practical eigenvalue in the circuit is termed $\lambda_G$. At the steady state, assuming the output voltages of inverters constitute a column vector $\boldsymbol{v}$, the crosspoint MVM currents are $\boldsymbol{i} = \boldsymbol{G_A}\boldsymbol{v}$, thanks to the virtual ground of the inverting-input node of TIAs. TIAs convert the crosspoint MVM current current into voltages, namely $\boldsymbol{v}_{TIA} = -\boldsymbol{G_A}\boldsymbol{v}/G_\lambda$. $\boldsymbol{v}_{TIA}$ is inverted by analog inverters, and the resulting voltages should be equal to $\boldsymbol{v}$, that is $\boldsymbol{G_A}\boldsymbol{v}/G_\lambda = \boldsymbol{v}$, or $\boldsymbol{G_A}\boldsymbol{v} = G_\lambda\boldsymbol{v}$. To satisfy this equation, $\boldsymbol{v}$ must be an analogous solution to Equation 1. As a result, Equation 1 is physically solved in one step by the circuit where $\boldsymbol{v}$ represents the eigenvector. Note that $\lambda$ is real and positive in Figure 1a, which is always the case for the largest eigenvalue of a positive matrix, according to the Perron-Frobenius theorem.[13] For the negative $\lambda$ case, the inverters in the circuit should be removed, and the absolute value of $\lambda$ is mapped by the feedback conductance $G_\lambda$.[8] In Figure 1a we consider a positive matrix, since the conductance of a resistive memory device can only be positive. For matrices containing negative elements, two crosspoint arrays are needed to split the matrix with two positive matrices.[8]

The in-memory calculation of eigenvectors was conducted in an array of resistive switching memory (RRAM) devices. In the RRAM device, the conductance can be changed by the formation and the dissolution of a conductive filament by local migration of ionized defects.[14] The RRAM conductance can be continuously tuned, thus enabling the analog storage in a crosspoint array for in-memory matrix computation.[5,8] Figure 1b shows 12 conductance levels of the adopted Ti/HfO$_x$/C RRAM device by controlling the compliance current during the set transition. As an experimental demonstration, the dominant eigenvector of a 3×3 positive matrix was computed by the circuit, with the programmed conductance matrix of a crosspoint array shown in the inset of Figure 1a. The experimental eigenvector results are shown in Figure 1c as a function of the normalized analytical eigenvector obtained by software calculations with floating-point precision. The linear relation between the two solutions demonstrates the circuit functionality of one-step eigenvector computation. By defining the solution error as $\epsilon = \|x - x^*\|$, where $x$ and $x^*$ are the experimental and the ideal normalized eigenvector, respectively, and $\|\cdot\|$ is the Euclidean norm, an error $\epsilon = 0.0303$ is found in Figure 1c. The 12 discrete conductance levels in Figure 1b were used to build matrices of various size for simulating the eigenvector circuit. For instance, the dominant eigenvectors of 10 randomly-constructed 10×10 matrices were computed in SPICE (simulation program with integrated circuit emphasis) circuit. In all cases the results are highly consistent with the analytical solutions, with an average error $\epsilon = 0.0056$ (Figure S1, Supporting Information).

In the conventional power iteration method, MVM is executed through element-wise multiply-accumulate operations and a number of iterations are required, resulting in a high computational complexity.[12,15,16] On the other hand, the MVM is instantaneously executed in the eigenvector circuit by physical laws in the crosspoint array, while discrete iterations are eliminated in favor of a higher computational speed.

To analyze the time complexity, the eigenvector circuit is illustrated as a block diagram (**Figure 2**a), where $x$ is the eigenvector solution mapped by the output voltages of the inverters and $y$ represents the output voltages of TIAs. Due to the crosspoint RRAM array acting as a feedback network, the transfer function linking $x$ to $y$ should be a matrix that involves the stored coefficient matrix $A$, the mapped eigenvalue $\lambda_G$ and the open-loop gain $L(s)$ of the operational amplifiers (OAs) which is a function of the complex frequency $s$. The transfer function linking $y$ backwards to $x$ is a scalar that is related solely to $L(s)$. Specifically, according to the Kirchhoff's voltage law and amplifier theory, $x$ and $y$ satisfy the following two equations:

$$-U[Ax(s) + \lambda_G y(s)]L(s) = y(s), \tag{2-1}$$

$$-\frac{x(s)+y(s)}{2}L(s) = x(s), \tag{2-2}$$

where $A$ is the $N \times N$ coefficient matrix and $U$ is a diagonal matrix defined as $U = diag\left(\frac{1}{\lambda_G+\sum_i A_{1i}}, \frac{1}{\lambda_G+\sum_i A_{2i}}, \cdots, \frac{1}{\lambda_G+\sum_i A_{Ni}}\right)$. We assumed that the OAs in both the TIAs and the inverters are identical, thus the same $L(s)$ applies to all the OAs. Combining the two equations of Equation 2 leads to:

$$U(A - \lambda_G I)L(s)x(s) = (2\lambda_G U + I)x(s) + \frac{2x(s)}{L(s)}, \tag{3}$$

where $I$ is the $N \times N$ identity matrix. Considering the single-pole OA model,[17] namely $L(s) = \frac{L_0}{1+\frac{s}{\omega_0}}$, where $L_0$ is the DC open loop gain and $\omega_0$ is the 3-dB bandwidth, Equation 3 becomes:

$$U(A - \lambda_G I)x(s) = \frac{2\lambda_G U + I}{L_0 \omega_0} s x(s) + \frac{2}{L_0^2 \omega_0^2} s^2 x(s), \tag{4}$$

where the insignificant terms have been omitted, due to the fact that $L_0$ is usually much larger than 1. The inverse Laplace transform of Equation 4 implies a second-order differential equation in the time domain, that is:

$$\frac{d^2 x(t)}{dt^2} = \frac{1}{2}L_0^2 \omega_0^2 U(A - \lambda_G I)x(t) - \left(\lambda_G U + \frac{1}{2}I\right)L_0 \omega_0 \frac{dx(t)}{dt}, \tag{5}$$

which describes the time response of the eigenvector circuit. To study the computing time of the circuit, Equation 5 is converted into a first-order differential equation[18] by defining:

$$z(t) = \frac{2}{L_0\omega_0}\frac{dx(t)}{dt}, \tag{6-1}$$

which leads to:

$$\frac{dz(t)}{dt} = L_0\omega_0 U(A - \lambda_G I)x(t) - L_0\omega_0\left(\lambda_G U + \frac{1}{2}I\right)z(t). \tag{6-2}$$

The two equations of Equation 6 are merged as one, which reads:

$$\frac{d}{dt}\begin{bmatrix} x(t) \\ z(t) \end{bmatrix} = L_0\omega_0 \begin{bmatrix} 0 & \frac{1}{2}I \\ U(A - \lambda_G I) & -\left(\lambda_G U + \frac{1}{2}I\right) \end{bmatrix}\begin{bmatrix} x(t) \\ z(t) \end{bmatrix}, \tag{7}$$

where $\mathbf{0}$ is the $N \times N$ zero matrix. By defining the $2N \times 2N$ matrix $\mathbf{M}$ according to:

$$M = \begin{bmatrix} 0 & \frac{1}{2}I \\ U(A - \lambda_G I) & -\left(\lambda_G U + \frac{1}{2}I\right) \end{bmatrix}, \tag{8}$$

which is associated with matrix $\mathbf{A}$, and defining a $2N \times 1$ vector as $\mathbf{w}(t) = \begin{bmatrix} x(t) \\ z(t) \end{bmatrix}$, Equation 7 becomes:

$$\frac{d\mathbf{w}(t)}{dt} = L_0\omega_0 \mathbf{M}\mathbf{w}(t). \tag{9}$$

According to the finite difference (FD) method, Equation 9 can be expressed as:

$$\mathbf{w}(t + \Delta t) = (\mathbf{I}_{2N} + \alpha\mathbf{M})\mathbf{w}(t), \tag{10}$$

where $\mathbf{I}_{2N}$ is the $2N \times 2N$ identity matrix, $\Delta t$ is the incremental time and $\alpha$ is a dimensionless constant defined as $\alpha = L_0\omega_0\Delta t$.

For Equation 9 to have a nontrivial solution, the spectral radius of matrix $\mathbf{I}_{2N} + \alpha\mathbf{M}$ has to be larger than 1, which implies that the highest eigenvalue (or real part of eigenvalue) $\lambda_h$ of matrix $\mathbf{M}$ must be positive, assuming the eigenvalues of $\mathbf{M}$ are ranked in a descending order according to their real parts. This condition on $\lambda_h$ is satisfied if the implemented $\lambda_G$ is slightly smaller than the largest eigenvalue of $\mathbf{A}$, namely $\lambda_G = (1 - \delta)\lambda_{max}$, where $\delta$ is a small positive number and $\lambda_{max}$ is the largest eigenvalue, thus the dominant eigenvector of $\lambda_{max}$ can be computed by the circuit. According to the FD algorithm of Equation 10, the eigenvector solution is boosted as long as the circuit is powered, until the boundary condition is encountered, *i.e.*, upon reaching the supply voltage $V_{supp}$ of the OAs. The parasitic noise in the circuit provides the initial solution $x(0) \neq 0$ at $t = 0$ which initiates the transient evolution of the circuit in Equation 10. At the steady state, the first $N$ elements of $w$ constitute the dominant eigenvector, while the last $N$ elements that represent the time derivative of $x(t)$ are zero.

To assess the circuit dynamics, we simulated the time evolution of $x(t)$ from the FD model in Equation 10 for the experimental matrix in Figure 1c. Figure 2b shows the simulation results, where we assumed $\delta = 0.01$ and $V_{supp} = \pm 1$ V for generality. A lower $V_{supp}$ or a diode connected to the output of the OAs might be used to protect the crosspoint memory devices against a possible voltage disturb if necessary. The results of the FD simulations are fully consistent with the SPICE circuit simulation results, demonstrating a good description of the circuit dynamics by Equation 10. The half-logarithmic plot in the inset evidences the exponential increase of the output voltages which can be explained by a power iteration where the output voltage is regenerated at each cycle in the closed-loop feedback circuit. The physical power iteration process stops at the saturation of the largest output voltage, after which point all other output voltages quickly stabilize and provide the final eigenvector solution.

According to Equation 10, the speed of the eigenvector circuit is controlled by $\lambda_h$ of matrix $M$, namely, the larger the $\lambda_h$, the faster the computation. To study the computing time of the circuit, we conducted a series of simulations by varying the eigenvalue difference $\delta$ for the eigenvector computation in Figure 1c. **Figure 3** shows the computed saturation time as a function of $\delta$ in the range from 0.003 to 0.06. In the simulations, the initial solution of each output was assumed as $x_i(0) = 0.001$. The computing time is defined as the time at which point the dynamic solution approaches the steady-state solution with an error less than 0.1%. It decreases with the inverse of $\delta$, with the time being 15.2 $\mu$s for the case of $\delta = 0.06$. The $\lambda_h$ for each case is also summarized in the figure, which shows a linear dependence on $\delta$, thus supporting the dominant role of $\lambda_h$ in controlling the computing time of the circuit. Though the circuit gets faster for a larger $\delta$, the resulting eigenvector deviates more from the ideal solution, as shown in the inset of Figure 3. The solution error $\epsilon$ increasing with $\delta$ is shown in Figure S2 (Supporting Information). These results indicate a tradeoff between the computing speed and the solution accuracy, which should be addressed according to the specific application scenario.

To study the dependence of computing time on matrix size *N*, we constructed a series of random matrices using the 12 conductance levels in Figure 1b. The size ranges from 3×3 to 30×30, as shown in **Figure 4**a with 3 representative matrices. For each size, 100 matrices were randomly generated and simulated in the circuit. The dominant eigenvector of every matrix was computed in simulation by assuming different $\delta$'s, namely $\delta = 0.003, 0.01, 0.02, 0.04$. Figure 4b shows the computing time as a function of *N*: the computing time is independent of *N*, thus demonstrating the $O(1)$ time complexity of the circuit for eigenvector computation. The computing time decreases for increasing $\delta$, in agreement with the results in Figure 3. For the 100 different matrices with the identical *N* and $\delta$, the computing time distribution is very tight, which further supports the dominant role of $\delta$ in controlling the computing time. Note that $\lambda_h$ increases with $\delta$, which is also consistent with the results in Figure 3. The solution errors are also independent of *N* (see Figure S3 of the Supporting Information), thus supporting the scalability of time/energy efficiency gain combined with no accuracy loss.

One concern about the computing time analysis is the parasitic wire resistance in the crosspoint array.[19] To investigate the impact of wire resistance on the time complexity of the circuit, we considered the interconnect parameters at 65 nm adopted from the ITRS (International Technology Roadmap for Semiconductors) table.[20] For the same dataset in Figure 4, the circuit simulations including wire resistances were conducted. The computing time for matrices with different sizes and $\delta$'s are shown in Figure S4 (Supporting Information). Compared to the ideal circuit, the real circuit including parasitic resistances generally shows a longer computing time, which may show a legible *N*-dependence. On the other hand, the solution error becomes lower for the real circuit, suggesting that wire resistances equivalently reduce the mismatch degree $\delta$ of eigenvalue implementation in the circuit. These results indicate an additional constraint on the tradeoff between the computing time and solution accuracy imposed by the wire resistance issue.

As a practical case study, we addressed the PageRank of a real-world dataset. The PageRank algorithm is widely used for ranking webpages in search engines,[9] link prediction and recommendation in social media.[21] PageRank aims at calculating the dominant eigenvector,[22] which can be naturally accelerated by the crosspoint eigenvector circuit. We adopted the Harvard500 database,[23] which contains 500 relevant webpages of the Harvard University to be ranked according to their connections. In the PageRank of a webpage network, the citations among webpages give a citation matrix $C$, which is defined as follows: if page *j* contains a link to page *i*, the citation element $C_{ij}$ is set to 1, otherwise $C_{ij} = 0$. More pages citing the same page indicates that the latter is more important. Also, citation by important pages gives rise to the importance of the page. **Figure 5**a shows the citation matrix of Harvard500, which is a sparse logical matrix. To rank the webpages by their importances, a transition matrix $T$ is defined according to:

$$T_{ij} = \begin{cases} \frac{pC_{ij}}{\sum_i C_{ij}} + \sigma, & if \sum_i C_{ij} \neq 0, \\ 1/N, & if \sum_i C_{ij} = 0. \end{cases} \qquad (11)$$

where $N = 500$ is the number of pages, $p = 0.85$ is the random walk probability, $\sigma = \frac{1-p}{N}$ is the probability for randomly picking a page. A uniform probability $1/N$ is assigned if a page gets no link.[23] The transition matrix is basically a stochastic matrix with the largest eigenvalue always being 1 and the dominant eigenvector giving the importance scores of webpages.[22] The resulting transition matrix for Harvard500 is illustrated in Figure S5 (Supporting Information).

The transition matrix was stored in the crosspoint array, and the largest eigenvalue was mapped in the feedback conductance with a mismatch degree $\delta$ to compute the eigenvector in the circuit. Figure 5b shows the circuit dynamics of the eigenvector computation with $\delta = 0.01$, indicating a computing time of around 135 $\mu$s. The simulated eigenvector values were normalized so that the sum of all elements is equal to one, thus giving the importance scores of the webpages. The results are shown in Figure 5c, where page #1 (the home page of Harvard university) ranks in the first place in the search results.

To study the time complexity of PageRank for webpage networks with different sizes, we selected the first $N$ pages in the Harvard500 database to form a new network, for which a new set of importance scores is computed. $N = 4, 8, 16, 32, 64, 128, 256$ were assumed, and the $N \times N$ submatrix was extracted from the entire citation matrix, as illustrated in Figure 5d. For each subset of webpages, $\delta = 0.003, 0.01, 0.02, 0.04$ were assumed for the eigenvalue implementation in the circuit. The computing time of all simulation cases are reported in Figure 5e, which also includes the computing time for the original Harvard500 with the four $\delta$'s. For a specific $\delta$, the computing time for all matrices are approximately at the same level, whereas a different $\delta$ causes a significant change of computing time. The scattering of the computing time over the $N$ range is due to the variation of $\lambda_h$ with $N$ (Figure S6, Supporting Information), which in turn is due to the specific structure of the transition matrices. These results support the $O(1)$ time complexity of the eigenvector circuit for practical application of PageRank.

Regarding the solution accuracy of PageRank, we show the comparison between the simulated importance scores and the ideal ones for the Harvard500 database in Figure S7 (Supporting Information), indicating a good consistency between the two solutions. In particular, we ranked the top 10 pages for the ideal case and the 4 simulated cases, showing that all the ideal top 10 pages are preserved in the top 10 places in the simulations, except for the case with $\delta = 0.04$, where one page was missed out. We also studied the wire resistance issue for the PageRank of Harvard500 subsets, with the results shown in Figure S8 (Supporting Information). The parasitic wire resistance causes a small increase of computing time for relatively large $\delta$, thus leading to an *N*-dependence to the time complexity of eigenvector computation. These results suggest a careful choice of $\delta$ for circuit implementation to achieve the best performance regarding both the computing time and the accuracy of the results. A strategy of dynamic tuning of $\delta$ might be adopted to achieve both high speed and accuracy. In this algorithm, a large $\delta$ can be used in the initial phases to accelerate the transition of the output voltages, then $\delta$ can be reduced in the later stages for fine tuning of the final solution.

As the mismatch degree $\delta$ is generally considered to be small to maintain the eigenvector accuracy, it may suffer from the conductance variation, *i.e.*, the feedback conductance values of the TIAs being slightly different. In this case, the associated matrix of Equation 8 becomes

$$M = \begin{bmatrix} \mathbf{0} & \frac{1}{2}I \\ U(A - \Lambda) & -\left(\Lambda U + \frac{1}{2}I\right) \end{bmatrix}, \tag{12}$$

where $\Lambda$ is a diagonal matrix that is defined as $\Lambda = diag\left(\lambda_G^{(1)}, \lambda_G^{(2)}, \cdots, \lambda_G^{(N)}\right)$, assuming $\lambda_G^{(i)}$ is the *i*-th practical implementation of the nominal eigenvalue $\lambda$. We simulated the PageRank of Harvard500 in the circuit by considering the eigenvalue variations. Instead of a uniform $\delta = 0.01$ in Figure 5b, the $\delta$ for each eigenvalue conductance was assumed lying randomly in the range of (0, 0.02). 10 trials were tested. All the results of computing time and solution error show a tight distribution around the ones of the uniform case (Figure S9, Supporting

Information), thus confirming the robustness of the circuit against feedback conductance variations in a practical implementation.

In conclusion, we have studied the time response of the crosspoint RRAM circuit for eigenvector computation, based on the feedback analysis of the self-sustained system. The circuit shows a computing time that relies solely on the mismatch degree of eigenvalue implementation in the circuit, which governs the convergence rate of output voltages toward the steady state solution. The computing time shows no dependence on the matrix size *N*, which supports the $O(1)$ complexity of the crosspoint eigenvector circuit. The PageRank of the Harvard500 database and its subsets also supports the $O(1)$ time complexity of the circuit. With such a low time complexity, this work supports the significant time/energy efficiency gains of in-memory computing for big data analytics in a wide range of real-world applications.

**Experimental Section**

*Experimental Devices*: The RRAM devices characterized in this work employ an $HfO_2$ thin film as the switching layer, whose thickness is 5 nm. The $HfO_2$ dielectric layer was deposited by e-beam evaporation on a confined graphitic carbon bottom electrode (BE), then a Ti thin layer was deposited on top of the dielectric layer as top electrode (TE) without breaking the vacuum during evaporation. The forming process of RRAM was operated by applying a DC voltage sweep from 0 to 5 V to TE with BE being grounded. The forming process induces a soft breakdown of the dielectric $HfO_2$ layer, initiating the CF formation and the resistive switching behavior. The set and reset transitions take place under positive and negative voltages applied to the TE, respectively. The DC conduction and switching characteristics of the RRAM were collected by a Keysight B1500A Semiconductor Parameter Analyzer, which was connected to the RRAM device in a conventional probe station for electrical characterization.

*SPICE Simulations*: Simulations of the crosspoint circuit were carried out using LTSPICE (https://www.linear.com/solutions/1066). Linear resistors were employed to dictate the conductance values of the programmed RRAM levels, thus mapping a matrix in the crosspoint array. AD823 from Analog Devices was employed as the operational amplifier in the circuit, parameters can be found here (https://www.analog.com/en/products/ad823.html).

**Acknowledgements**
This article has received funding from the European Research Council (ERC) under the European Union's Horizon 2020 research and innovation programme (grant agreement No 648635).

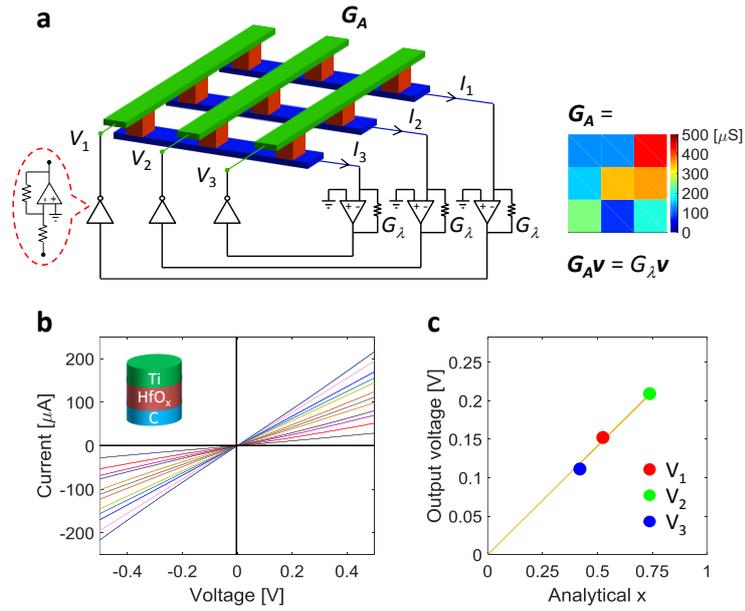

**Fig. 1.**

**Figure 1.** Eigenvector computation with a crosspoint RRAM circuit. a) The crosspoint RRAM circuit for computing the dominant eigenvector of a positive matrix. The circuit structure of the analog inverter is also shown. The output voltages of analog inverters form a vector $\boldsymbol{v} = [V_1; V_2; V_3]$ representing the eigenvector solution. The crosspoint MVM currents form $\boldsymbol{i} = [I_1; I_2; I_3]$. A representative matrix is also shown, with a conductance unit of 100 $\mu S$. b) 12 analog conductance levels of the Ti/HfO$_x$/C RRAM device, with values of 60, 90, 120, 150, 190, 210, 240, 290, 310, 340, 390 and 420 $\mu S$, respectively. The conductance shows a good linearity for all levels below 0.5 V. c) The dominant eigenvector of the matrix computed by the circuit, as a funtion of the analytical solution.

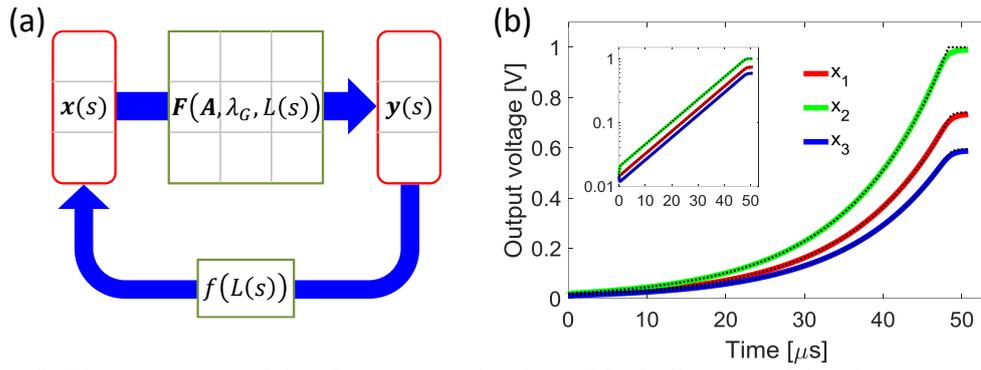

**Figure 2.** Time response of the eigenvector circuit. a) Block diagram of the eigenvector circuit, where $x$ and $y$ represent the output voltages of analog inverters and TIAs, respectively. The transfer function $F$ from $x$ to $y$ is a matrix, while the transfer function $f$ feeding $y$ back to $x$ is a scalar. b) Transient simulation results of computing the eigenvector in Figure 1c. The color full lines are SPICE simulation traces, while the dot lines are the iterative simulation result according to the FD algorithm. The inset shows the same results on a half-logarithmic plot to highlight the exponential increase of the output voltages.

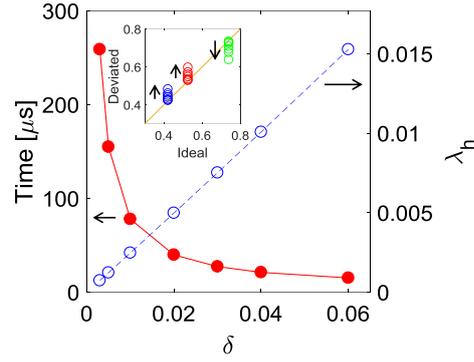

**Figure 3.** Analysis of $\lambda_h$ of the associated matrix ***M*** and computing time of the circuit. Various mismatch degree $\delta$'s were introduced in the eigenvalue implementation in circuit simulations. The computing time is proportional to $1/\delta$, while the $\lambda_h$ is proportional to $\delta$. The inset shows the correlation plot between the simulated eigenvectors and the ideal eigenvector, with the arrows indicating the increase of $\delta$.

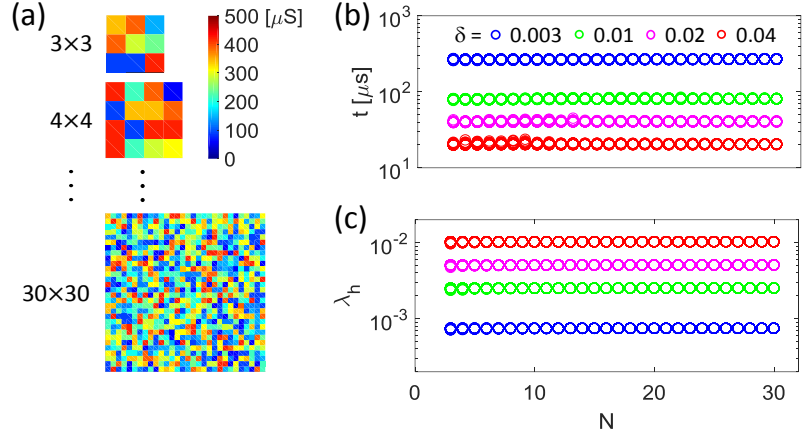

**Figure 4.** Time complexity for computing dominant eigenvectors of random matrices. a) Illustrations of random matrices with sizes from 3×3 to 30×30. 100 matrices were constructed with the 12 discrete conductance levels in Figure 1b for each size. b) Computing time of the dominant eigenvectors of the matrices with different size, with 4 different $\delta$'s introduced in the eigenvalue implementation. c) $\lambda_h$'s of all simulation cases. Both the computing time and $\lambda_h$ show a clear independence on the matrix size $N$, indicating the $O(1)$ time complexity.

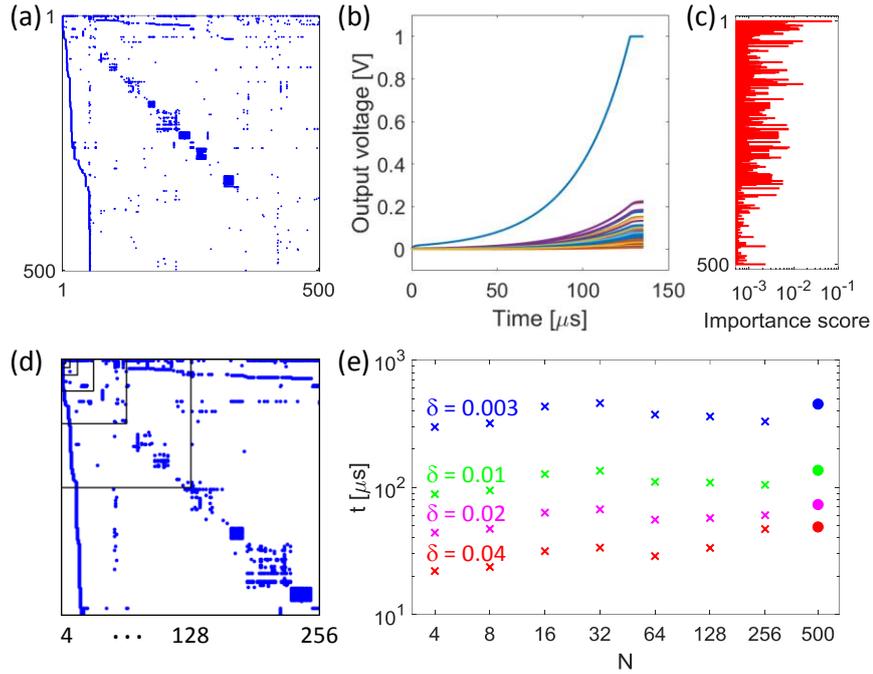

**Figure 5.** Time complexity of PageRank with the Harvard500 database and its subsets. a) The citation matrix of Harvard500, which is a sparse matrix containing 2,636 connections among webpages in the entire database. b) Transient simulation results of PageRank for Harvard500 assuming $\delta = 0.01$. c) Importance scores of the 500 webpages obtained from the simulated eigenvector. d) Citation matrices of subsets of the Harvard500 database, with different sizes from 4×4 to 256×256. e) Computing time of PageRank of Harvard500 and its subsets, with 4 different $\delta$'s assumed. For each $\delta$, the computing time of PageRank of various datasets are on the same level, thus supporting the $O(1)$ of the crosspoint circuit for eigenvector computation.

*Supporting Information of*

**In-memory eigenvector computation in time *O*(1)**

*Zhong Sun\*, Giacomo Pedretti, Elia Ambrosi, Alessandro Bricalli, and Daniele Ielmini\**

Dipartimento di Elettronica, Informazione e Bioingegneria, Politecnico di Milano and IU.NET, Piazza L. da Vinci 32 – 20133 Milano, Italy
E-mail: zhong.sun@polimi.it; daniele.ielmini@polimi.it

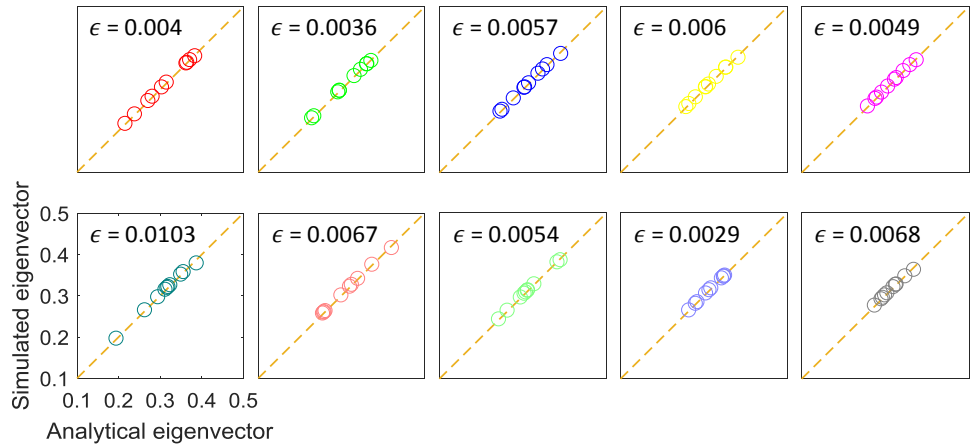

**Figure S1.** Correlation plots of the dominant eigenvectors computed by the crosspoint RRAM circuit and their ideal analytical solutions. Ten 10×10 positive matrices were randomly constructed with the 12 discrete conductance levels in Figure 1b. The circuit was simulated in SPICE. Both sets of solutions show a high consistency for all simulation cases, with the solution error $\epsilon$ identified for each case.

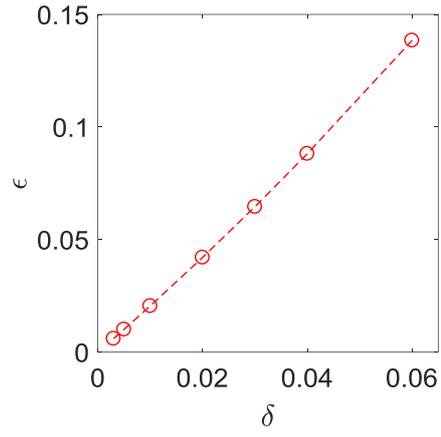

**Figure S2.** Solution errors of eigenvector computation in the circuit with different mismatch degrees of eigenvalue implementation.

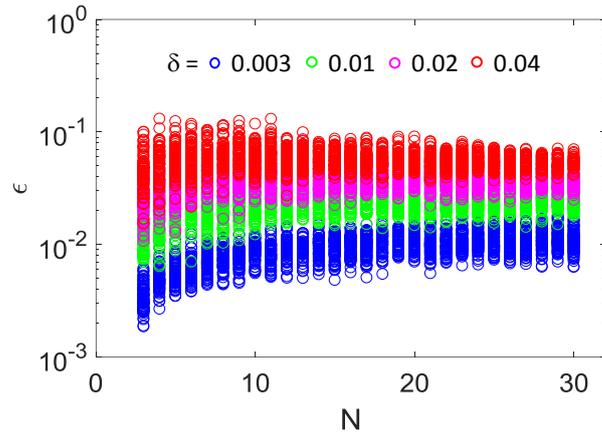

**Figure S3.** Solution errors of the simulated eigenvectors of random matrices. The error $\epsilon$ remains constant as $N$ increases.

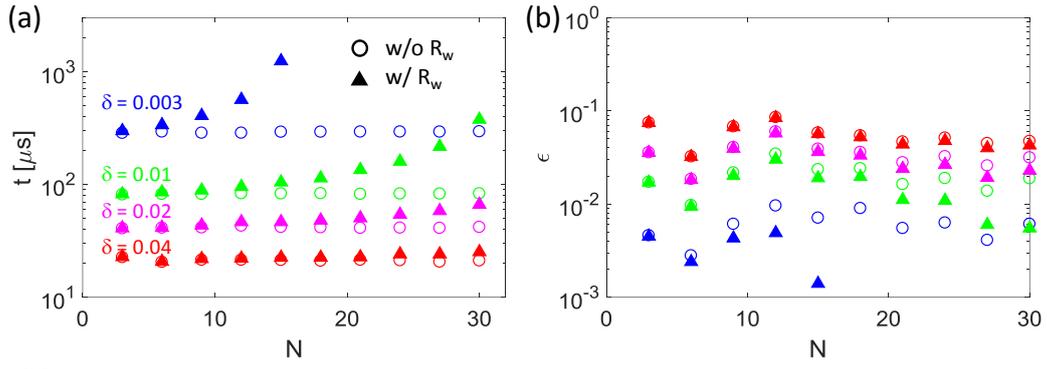

**Figure S4.** a) Computing time of the real circuit with wire resistances. The matrix dataset is the one in Figure 4, but only specific sizes were considered, namely $N$ = 3, 6, 9 …, 30. Only one matrix was simulated for each size. The circuit without wire resistances shows the $O(1)$ time complexity. The parasitic wire resistances causes a weak $N$-dependence of the time complexity, which becomes more obvious for a small $\delta$, *e.g.* $\delta = 0.003$. b) Solution error of the real circuit. The $N$-dependence of solution error of the real circuit shows the opposite behavior of the computing time.

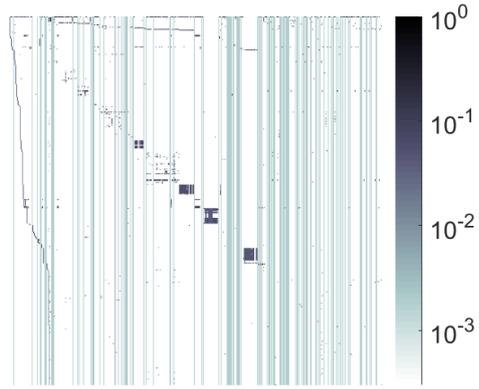

**Figure S5.** The transition matrix of Harvard500. As the citation matrix is sparse, the transition matrix contains mostly small values, for instance, 74.55% of the entries are $3 \times 10^{-4}$, and 24.4% of the entries are $2 \times 10^{-3}$.

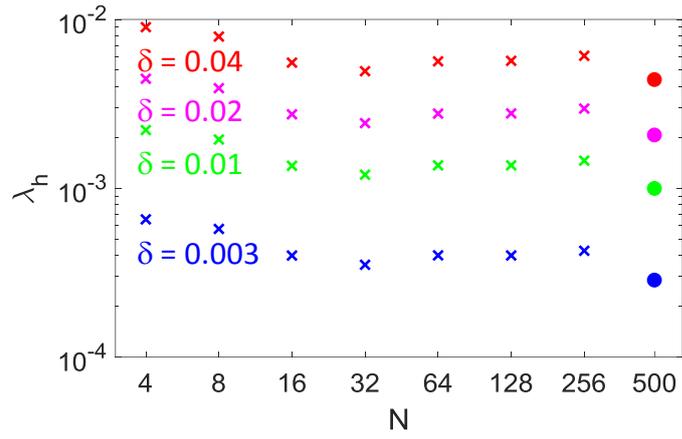

**Figure S6.** $\lambda_h$ in the simulation cases of Harvard500 and its subsets with different $\delta$'s assumed in the circuit implementation.

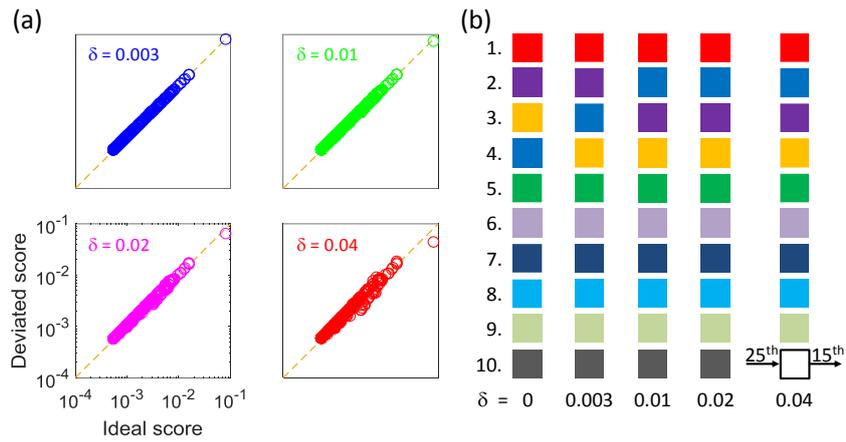

**Figure S7.** PageRank results of Harvard500 with different $\delta$'s. a) Correlation plots of the simulated importance scores as a function of the ideal ranking result. Although the accuracy of simulation result decreases as $\delta$ increases, it is considerably high for relatively small $\delta$'s. b) The top 10 pages in the ranking results for the ideal case and the simulation cases with different $\delta$'s. The $\delta = 0$ case shows the ideal top 10 webpages, with each color representing a certain website in Harvard500. For instance, the red symbol represents the home page of Harvard university, which is ranked in the first place. For the other simulation cases, the ranking may be incorrect, however all pages preserve their position in the first 10 places, except for the case of $\delta = 0.04$ where a page is missed out, namely the ideal $10^{th}$ page is ranked in the $15^{th}$ place, and replaced by the ideal $25^{th}$ page.

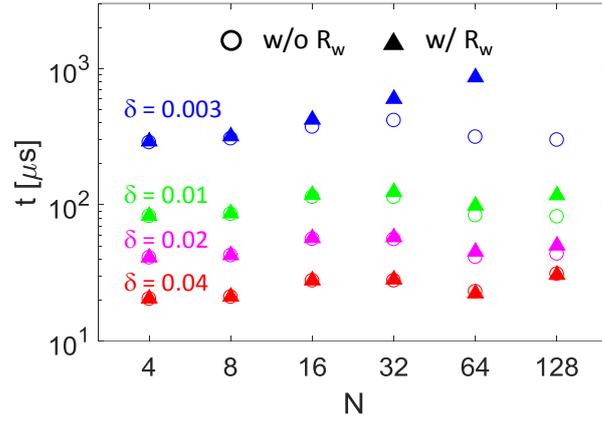

**Figure S8.** Computing time of PageRank simulations with wire resistances. The parasitic wire resistances cause a slight increase of the computing time for simulation cases of relatively large $\delta$'s. The time increment becomes significant for $\delta = 0.003$, resulting in an $N$-dependence of the time complexity of eigenvector computation.

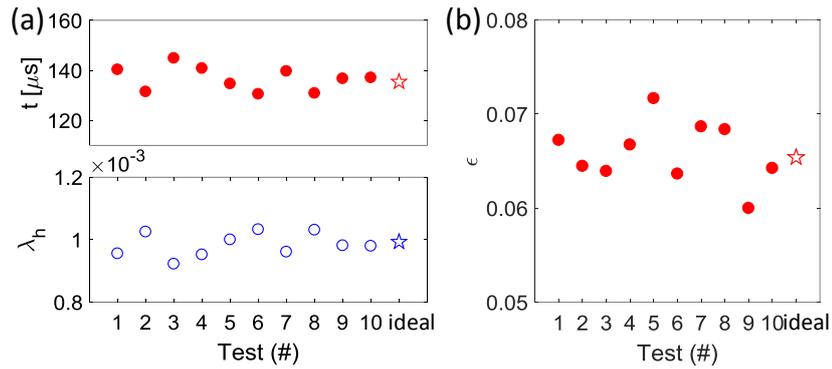

**Figure S9.** Eigenvector computation with feedback conductance variations. a) Computing time (top panel) and $\lambda_h$ (bottom panel), and b) Solution errors, of 10 random trials and the ideal case. In the ideal case, the eigenvalue implementation is assumed with a uniform $\delta = 0.01$ for all feedback conductance values. In the random trials, each feedback conductance value includes randomly a $\delta$ in the range of (0, 0.02).